\documentclass[sigconf]{acmart}

\usepackage{subcaption}

\copyrightyear{2023}
\acmYear{2023}
\setcopyright{rightsretained}
\acmConference[WWW '23 Companion]{Companion Proceedings of the ACM Web Conference 2023}{April 30-May 4, 2023}{Austin, TX, USA}
\acmBooktitle{Companion Proceedings of the ACM Web Conference 2023 (WWW '23 Companion), April 30-May 4, 2023, Austin, TX, USA}
\acmDOI{10.1145/3543873.3584641}
\acmISBN{978-1-4503-9419-2/23/04}

\begin{document}

\title{Latent User Intent Modeling for Sequential Recommenders}


\author{Bo Chang, Alexandros Karatzoglou, Yuyan Wang, Can Xu, Ed H. Chi, Minmin Chen}
\affiliation{
    Google, Inc. \\
    Mountain View, CA, USA \\
    \{bochang, alexkz, yuyanw, canxu, edchi, minminc\}@google.com
}

\renewcommand{\shortauthors}{Chang et al.}

\begin{abstract}
Sequential recommender models are essential components of modern industrial recommender systems. These models learn to predict the next items a user is likely to interact with based on his/her interaction history on the platform. Most sequential recommenders however lack a higher-level understanding of user intents, which often drive user behaviors online. Intent modeling is thus critical for understanding users and optimizing long-term user experience. We propose a probabilistic modeling approach and formulate user intent as latent variables, which are inferred based on user behavior signals using variational autoencoders (VAE). The recommendation policy is then adjusted accordingly given the inferred user intent. We demonstrate the effectiveness of the latent user intent modeling via offline analyses as well as live experiments on a large-scale industrial recommendation platform.
\end{abstract}




\maketitle

\section{Introduction}

Modern recommender systems are powering many online platforms. 
Sequential recommendation models that consider the order of user-item interactions are becoming increasingly popular~\citep{ijcai2019p883,chen2019top,tang2019towards,kang2018self}.
Most existing sequential recommender models mainly rely on the item-level interaction history of a user to capture his/her \emph{topical interests}. These models often lack a higher-level understanding of \emph{user intent}, i.e., what a user wants from the platform at request time, for example exploring new contents, continuing with contents from the last session, playing background music, etc. User intents often span across sessions, thus intent understanding is crucial to optimizing long-term user experience. 

User intent can be explicitly defined. For example, search intents are classified into navigational, informational, and transactional~\citep{broder2002taxonomy,rose2004understanding,jansen2007determining}.
With this approach, we can annotate the training data with explicitly defined user intent and cast it as a supervised learning task to predict user intent. 
This formulation has clear advantages: it allows for good interpretability and reliable evaluation of model prediction. 
It however also comes with an obvious downside, i.e., requiring expert knowledge to manually define and enumerate user intent. 
Compared with the search use case, user intent in organic recommendation is often multifaceted and subconscious, which are much more challenging to manually define~\citep{kalaganis2021unlocking}.

Rather than explicitly defining user intent, we propose to model user intent as latent variables using \emph{latent variable models}, which have been extensively studied and applied across various fields including statistics~\citep{everett2013introduction}, machine learning~\citep{bishop1998latent}, and econometrics~\citep{aigner1984latent}.
It is a probabilistic model relating observed variables or evidence to latent variables.
Specifically, it defines a joint distribution over the observed and latent variables. One can then obtain the corresponding distribution of the observed variables through marginalization.

User behavior signals on an online platform---for example, search, browse, click, and consumption---are often good indicators of the underlying user intent. 
Therefore, they can be used as observed variables in the latent variable model.
In other words, we formulate user intent as latent variables and jointly model them with user behavior signals, avoiding the need to manually define intent.

Together we make the following contributions:
\begin{itemize}
    \item Propose a probabilistic model to formulate the user intent as latent variables and relate it to behavior and context signals.
    \item Apply variational inference techniques for efficient and scalable inference.
    \item Shed light on training stability of probabilistic models for industrial use cases.
    \item Conduct an analysis to gain insights into the semantics of the latent space. 
    \item Demonstrate the benefits of the proposed technique in large-scale live experiments on a commercial recommendation platform serving billions of users and millions of items.
\end{itemize}

\section{Related Work}

\paragraph{Sequential recommendation systems}

Sequential recommenders are a class of recommender systems, which are based on modeling the order of the user-item interactions~\citep{ijcai2019p883}.
Various architectures have been applied to capture long-term and complex dependencies in user-item interactions, for example, recurrent neural networks (RNN)~\citep{Hidasi2016,beutel2018latent,wu2017recurrent,chen2019top}, convolutional neural networks (CNN)~\citep{tang2019towards,tang2018personalized,yuan2019simple}, and self-attentive models~\citep{kang2018self}. 
However, most methods mainly focus on the item-level interaction history and often lack a higher-level understanding of user intent.

\paragraph{User intent modeling}

Multiple approaches have been proposed to model user intent for improving recommender systems~\citep{li2021intention,pan2020intent,cen2020controllable,li2019multi,ma2020disentangled,tanjim2020attentive,chen2022intent};
in particular, the implicit user intent approach has gained popularity. 
\citet{wang2019modeling} use the mixture-channel purpose routing network (MCPRN) to learn users’ different purchase purposes of each item.
\citet{li2022automatically} propose methods to discover new intents based on existing explicitly-defined ones.
Implicitly deduced intent from user interactions has also been used to understand user satisfaction in music recommendation~\citep{spotify}.
As far as we know, our proposal to bring latent variable models into intent modeling and leverage behavior signals is a novel contribution.  

\paragraph{Latent variable models and variational autoencoders}

Latent variable models are a class of probabilistic models that assume the observed variables are generated by a set of unobserved or latent variables.
The models learn the joint distribution of observed and latent variables, then the distribution of observed ones is obtained by marginalization~\citep{everett2013introduction,bishop1998latent,aigner1984latent}.
Variational autoencoders (VAE) provide a principled framework for learning deep latent variable models and corresponding inference models~\citep{kingma2013auto,Kingma2019}. 
Several extensions of VAE are later proposed, including conditional VAE that models the conditional distributions~\citep{sohn2015learning} and variational RNN for modeling sequential data~\citep{chung2015recurrent}. 
Applications of VAE to recommender systems have also been an active research topic~\citep{liang2018variational,ma2018partial,luo2020deep,li2017collaborative}.
In this work, we propose a novel application of latent variable models for user intent modeling.

\section{Latent Intent Modeling}

In this section, we provide a detailed exposition of the proposed latent intent modeling technique.
Section~\ref{sec:intent_pgm} discusses the model assumptions in the form of a probabilistic graphical model. 
Section~\ref{sec:intent_vae} describes the model architecture and training objective of the latent intent module.
Finally, Section~\ref{sec:intent_overall} summarizes how the module is incorporated into a sequential recommender model.

\subsection{Probabilistic Model}
\label{sec:intent_pgm}

\begin{figure}[htbp]
    \centering
    \includegraphics[width=0.9\linewidth]{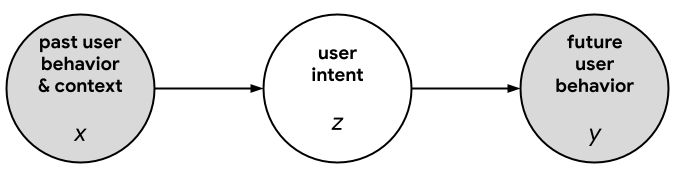}
    \caption{Probabilistic graphical model. The connection from past user behavior and context $x$ to future user behavior $y$ is mediated by the user intent $z$. Note that $x$ and $y$ are observed variables whereas $z$ is latent.}
    \label{fig:graphical_model}
\end{figure}

We start with a directed probabilistic graphical model (PGM) or Bayesian network to factorize the joint distribution of the random variables of interest.
The high-level structural assumption of the model is that past user behavior and context information before the user request indicates a user's current intent, which is further manifested in future user behavior.
Figure~\ref{fig:graphical_model} provides a graphical representation of the model. 
Following the convention of visual representations of probabilistic graphical models, we use shaded nodes to denote observed variables and transparent nodes for latent ones.
The nodes in the graph represent the following variables:
\begin{itemize}
    \item Node $x$: past user behavior (e.g., number of clicks/searches in the past 15 minutes) and context (e.g., time of day, device);
    \item Node $y$: future user behavior (e.g., number of clicks/searches in the next 15 minutes);
    \item Node $z$: latent variable (i.e., user intent).
\end{itemize}
Note that $x$ and $y$ are assumed to be conditionally independent given $z$. In other words, the link between past user behavior and context $x$ and future behavior $y$ are mediated by the latent intent $z$.

\subsection{Inference with Variational Autoencoders}
\label{sec:intent_vae}

From a probabilistic modeling perspective, the goal of the latent intent model is to capture the conditional distribution $p(y|x)$, where the link between $x$ and $y$ is mediated by the latent variable $z$. 
Latent variable models are known to be difficult to learn and infer from, due to intractable posterior distributions.
In order to make learning and inference tractable, we adopt a variational inference algorithm, i.e., the conditional VAE~\citep{kingma2013auto,sohn2015learning} to efficiently scale to large datasets.
The high-level idea is to introduce a variational distribution $q(z|x,y)$ that approximates the true posterior distribution and maximizes a lower bound of the log-likelihood: $\log p(y|x)$. It casts inference as an optimization problem. We refer interested readers to \citet{Kingma2019} and references therein for further details about variational inference and VAE.

The model is composed of the following networks/distributions:
\begin{itemize}
    \item Prior: $p(z|x)\sim\mathcal{N}(\mu_\psi, \Sigma_\psi)$;
    \item Decoder/likelihood: $p(y|z)\sim\mathcal{N}(\mu_\theta, \Sigma_\theta)$;
    \item Encoder/variational distribution: $q(z | x, y)\sim\mathcal{N}(\mu_\phi, \Sigma_\phi)$.
\end{itemize}
All distributions are parameterized as multivariate Gaussian with diagonal covariance matrices; the mean and log-variance are the output of multi-layer perceptrons (MLP), with ReLU as the activation function.

The parameters of the networks are trained to maximize the evidence lower bound (ELBO):
\begin{equation}
    \mathcal{L}_\text{ELBO} = \mathbb{E}_{q(z|x, y)} [ \log p(y | z) ] - D_\text{KL} ( q(z|x, y) \| p(z | x) ), 
    \label{eq:elbo}
\end{equation}
where $D_\text{KL}(\cdot \| \cdot)$ denotes the Kullback--Leibler (KL) divergence.
It can be shown that the ELBO is a lower bound of the conditional likelihood: 
$\mathcal{L}_\text{ELBO} \leq \log p ( y | x )$, which is the quantity we are interested in maximizing~\citep{sohn2015learning}.

The first term of the ELBO in Equation~\ref{eq:elbo} can be regarded as a \emph{reconstruction loss}. Given a pair of past user behavior and context $x$ and future user behavior $y$, if we pass it through the encoder and sample a latent intent $z$ from $q(z|x,y)$, then $z$ should contain relevant information about $y$ and the model will be able to truthfully reconstruct $y$.
The dimensionality of $z$ is often chosen to be smaller than that of $y$ so that it can act as an information bottleneck~\citep{alemi2017deep}.
The second term of the ELBO is a \emph{regularization loss}; it encourages the approximate posterior distribution $q(z|x,y)$ to not deviate too much from the prior distribution $p(z|x)$.

\begin{figure}[htbp]
    \centering
    \includegraphics[width=0.8\linewidth]{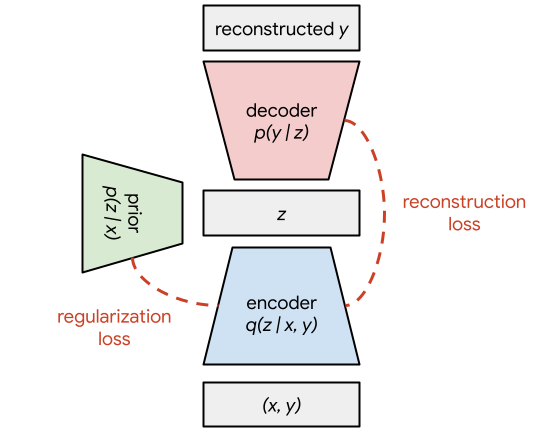}
    \caption{The latent intent module consists of three networks: prior, decoder; and encoder. The loss can be written as two terms: reconstruction loss and regularization loss.}
    \label{fig:vae_diagram}
\end{figure}

Figure~\ref{fig:vae_diagram} shows a diagram of the latent intent module.
Note that one adaptation from the original conditional VAE model is that the decoder does not take $x$ as input, i.e., $p(y|z)$ instead of $p(y|x,z)$. This is to be consistent with the probabilistic graphical model defined in Figure~\ref{fig:graphical_model}; that is, $y$ is conditionally independent of $x$ given $z$.

\paragraph{Training Stability} In practice, we observed training instability issues of the latent intent module, due to an exploding KL-divergence term, either at the very beginning or in the middle of model training.
Two techniques are found effective in mitigating the issues. 
The first one is the \emph{proper initialization} of the trainable parameters. 
We initialize the weights by a uniform distribution around zero $\mathcal{U}(-\varepsilon, \varepsilon)$, with $\varepsilon$ being a small positive number, and initialize the biases to zero.
As a result, both $q(z|x, y)$ and $p(z | x)$ are close to $\mathcal{N}(0, I)$ at initialization, thus the KL-divergence is near zero.
Secondly, \emph{soft-clipping the log-variance} throughout training can prevent KL-divergence from exploding~\citep{chua2018deep}. 
In particular, we use the softplus function to clip the log-variance to the interval $(a, b)$:
\begin{equation}
    f(x) = x - \log (1 + \exp (x-b) ) + \log (1 + \exp (a-x) ).
\end{equation}
Combining the two techniques together, we observe that the KL-divergence term remains bounded throughout model training.

\subsection{Incorporation into Sequential Recommenders}
\label{sec:intent_overall}

The latent intent model introduced in Section~\ref{sec:intent_vae} is a standalone module with its own training objective. 
In this section, we describe how it is incorporated into the main recommendation model, in particular, a REINFORCE sequential recommendation model~\citep{chen2019top}. Note that the proposed technique can be easily applied to other types of sequential recommendation models, for example, self-attentive ones~\citep{kang2018self}.

We briefly describe components of the recommendation model that are pertinent to this work and refer interested readers to \citet{chen2019top} for more details.
The sequential recommendation model uses an RNN to summarize a user's interaction history on the platform. 
The final RNN hidden state is concatenated with other context information and passed through the post-fusion layers (an MLP with ReLU activations), the output of which is used as the user representation. A softmax policy is defined on top of that, together with item representations. The model is trained using the REINFORCE algorithm~\citep{williams1992simple} to optimize long-term user satisfaction; we denote the loss function by $\mathcal{L}_\text{rec}$.

\begin{figure}[htbp]
    \centering
    \includegraphics[width=0.95\linewidth]{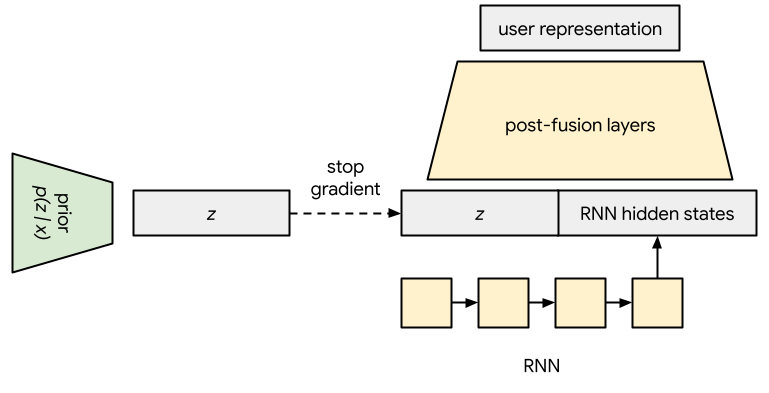}
    \caption{The overall architecture incorporating the latent intent module to the sequential recommendation model. Only the prior network is used and a stop gradient is applied to further decouple them.}
    \label{fig:overall_diagram}
\end{figure}

Among the three components of the latent intent model (prior, encoder, and decoder), we only make the prior network directly interact with the main recommender model. 
Recall that the prior network $p(z|x)$ infers user intent $z$ given past user behavior and context $x$. 
Both at training and serving time, we pass $x$ through the prior network and draw a sample $z$ from $p(z|x)$. 
In order to further decouple the latent intent module from the main recommendation model, we apply a stop gradient operation to $z$.
Then the sampled user intent $z$ is concatenated with the final RNN hidden state and then passed to the post-fusion layers. The output is the user representation augmented with the inferred user intent, which further conditions the recommendation policy.
The training objective of the sequential recommendation model $\mathcal{L}_\text{rec}$ remains unchanged. 
The overall loss function\footnote{Note that we want to maximize $\mathcal{L}_\text{ELBO}$, thus the minus sign before it in the loss function.} is 
$\mathcal{L} = \mathcal{L}_\text{rec} - \lambda \mathcal{L}_\text{ELBO}$,
where $\lambda > 0$ is a hyperparameter controlling the relative strength of the ELBO loss.
The overall model architecture is illustrated in Figure~\ref{fig:overall_diagram}.

\section{Latent Space Analysis}


\begin{figure*}[htbp]
    \centering
    \begin{subfigure}[t]{0.32\linewidth}
        \includegraphics[width=\textwidth]{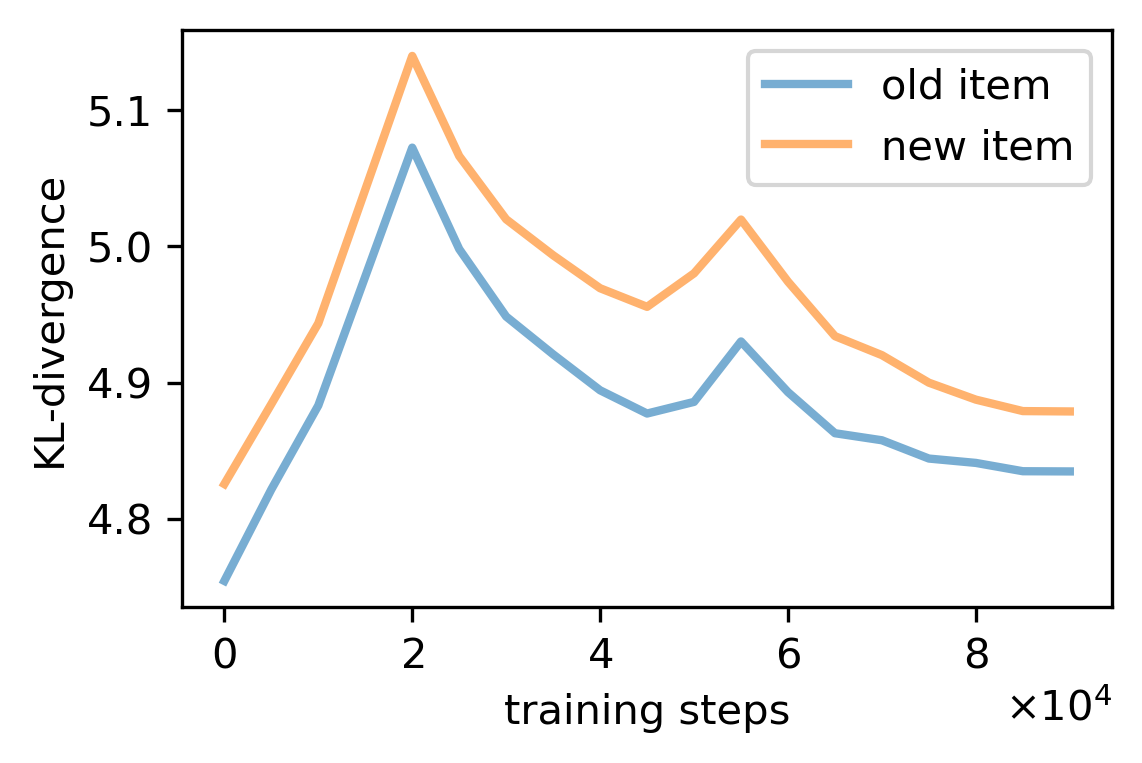}
        \caption{Item-level.}
        \label{fig:kl_analysis_item}
    \end{subfigure}
    \
    \begin{subfigure}[t]{0.32\linewidth}
        \includegraphics[width=\textwidth]{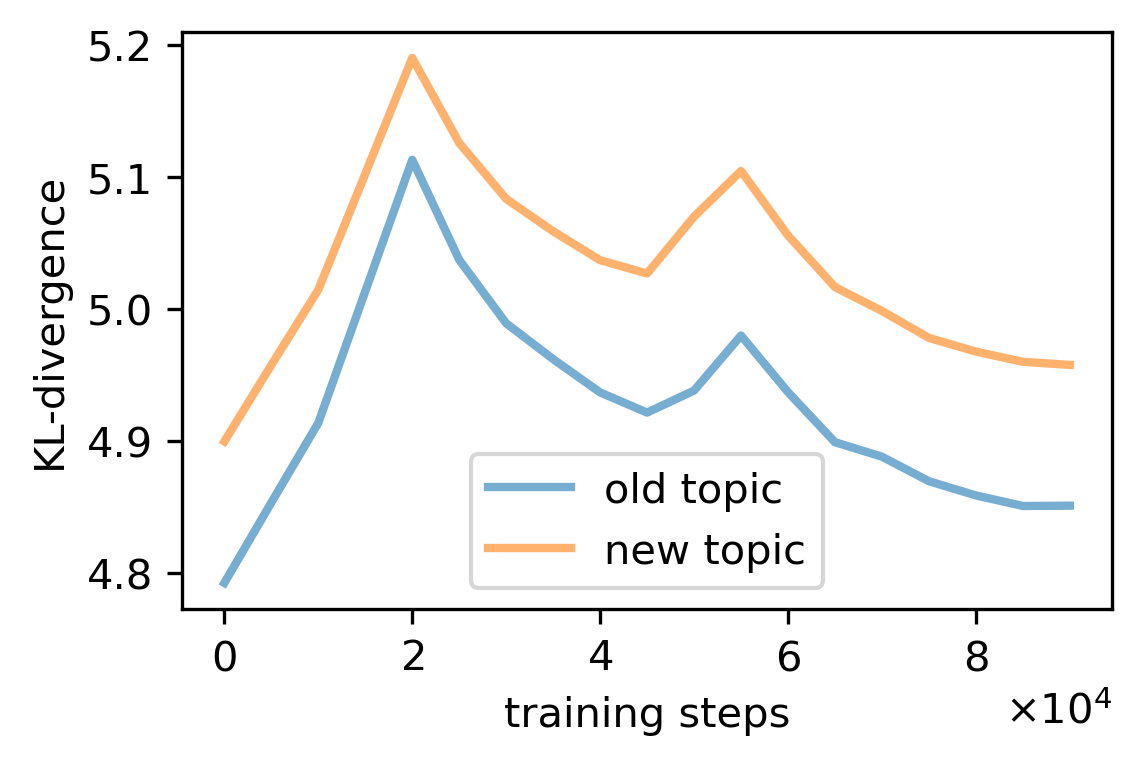}
        \caption{Topic-level.}
        \label{fig:kl_analysis_topic}
    \end{subfigure}
    \
    \begin{subfigure}[t]{0.32\linewidth}
        \includegraphics[width=\textwidth]{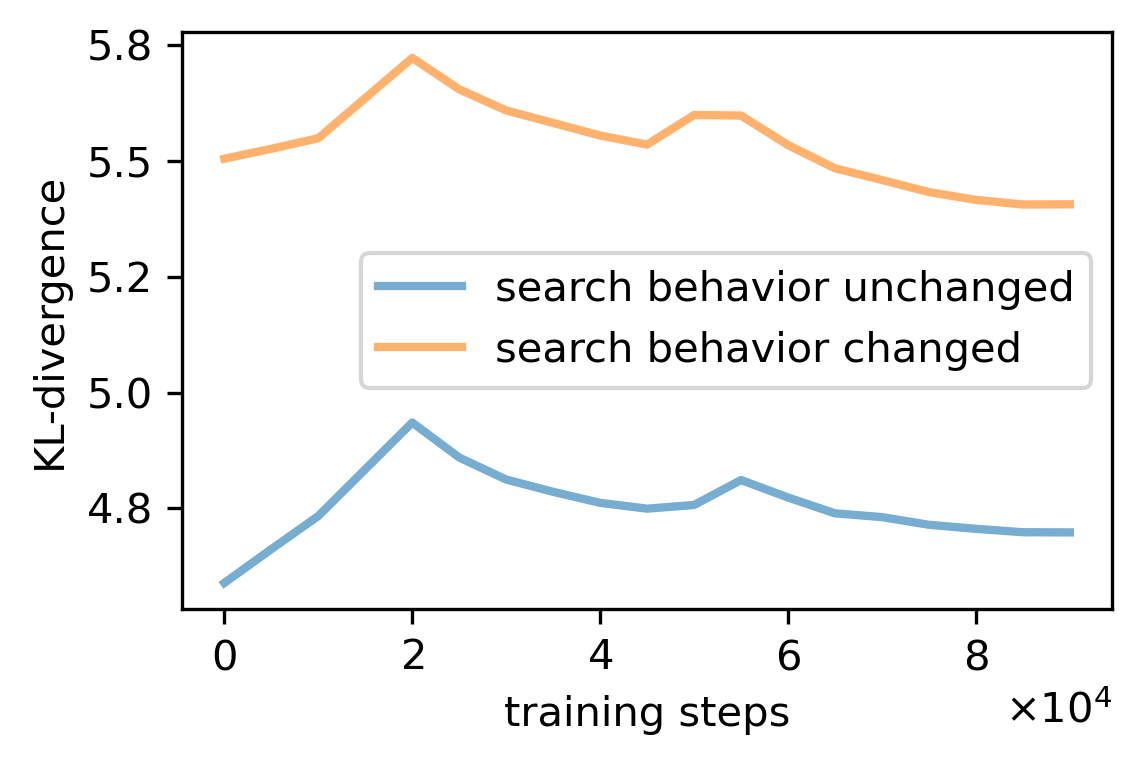}
        \caption{Search behavior.}
        \label{fig:kl_analysis_search}
    \end{subfigure}    
    \caption{Latent space analysis. The KL-divergence between the approximate posterior $q(z|x, y)$ and prior $p(z | x)$ is plotted against the training steps. The two curves correspond to new and old items at both item- and topic-levels in (a) and (b), and changed/unchanged search behavior in (c).}
    \label{fig:kl_analysis}
\end{figure*}


One drawback of VAE is the limited interpretability of the latent space.
In our use case, the inferred user intent $z$ is a compressed representation of user intent in a continuous latent space.
In order to gain more insights into what the latent space encodes, we adopt an analysis technique by \citet{chung2015recurrent}. 

Recall that the KL-divergence term in Equation~\ref{eq:elbo} is computed between the approximate posterior $q(z|x, y)$ and prior distributions $p(z | x)$; in other words, it can be seen as the difference in user intent with and without the knowledge of future user behavior.
Therefore, it can be regarded as a measure of \emph{surprise}.
In fact, the KL-divergence between posterior and prior distributions has long been used in neuroscience to quantify surprise~\citep{itti2005bayesian,friston2010free}.

Intuitively, when a user consumes an item that is ``new'', the amount of surprise and KL-divergence should be higher than that of an ``old'' item.
Therefore, we can compare the KL-divergence between new and old items consumed by users to corroborate the statement above.
Two definitions of new items are considered: (1) item-level: an item that has not been consumed by the user before; (2) topic-level: an item that belongs to a topic cluster\footnote{The topic clusters are produced by: 1) taking the item co-occurrence matrix, where the $(i,j)$-th entry counts the number of times item $i$ and $j$ were interacted by the same user consecutively; 2) performing matrix factorization to generate an embedding for each item; 3) using $k$-means to cluster the learned embeddings into 10,000 clusters; 4) assigning the nearest cluster to each item. } that has not been consumed by the user before.

We also consider scenarios when there are sudden changes in user search behavior, and the KL-divergence should be higher when such changes occur. 
In particular, we focus on the number of search queries a user issued/issues in the past 15 minutes $s_\text{past}$ and in the next 15 minutes $s_\text{future}$. 
We say there is a change in user search behavior if a user searched in the past but not in the future ($s_\text{past} > 0$ and $s_\text{future} = 0$) or vice versa ($s_\text{past} = 0$ and $s_\text{future} > 0$); otherwise, user search behavior is deemed unchanged. 

Figure~\ref{fig:kl_analysis} shows the average KL-divergence against the training steps.
Figures~\ref{fig:kl_analysis_item} and \ref{fig:kl_analysis_topic} compare ``new'' and ``old'' items at the item- and topic-levels, respectively.
When a user consumes a new item or topic, the KL-divergence is higher. 
Figure~\ref{fig:kl_analysis_search} shows the KL-divergence when user search behavior is changed/unchanged. 
When there is a change in user search behavior, the KL-divergence is higher.
These results indicate that the latent space is able to capture salient information and detect transitions in user behavior.

\section{Live Experiments}

\begin{figure}[htbp]
    \centering
    \begin{subfigure}[t]{0.49\linewidth}
        \includegraphics[width=\textwidth]{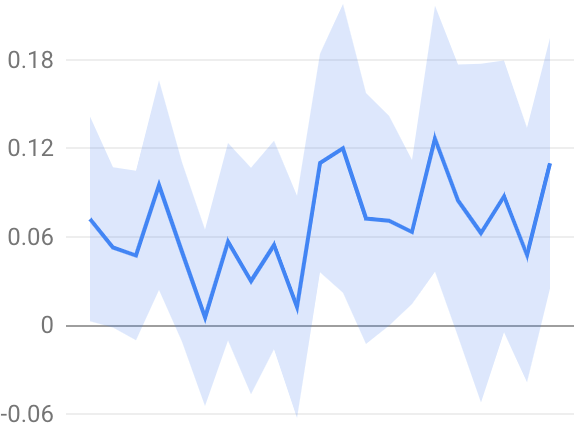}
        \caption{Overall enjoyment.}
    \end{subfigure}
    \
    \begin{subfigure}[t]{0.49\linewidth}
        \includegraphics[width=\textwidth]{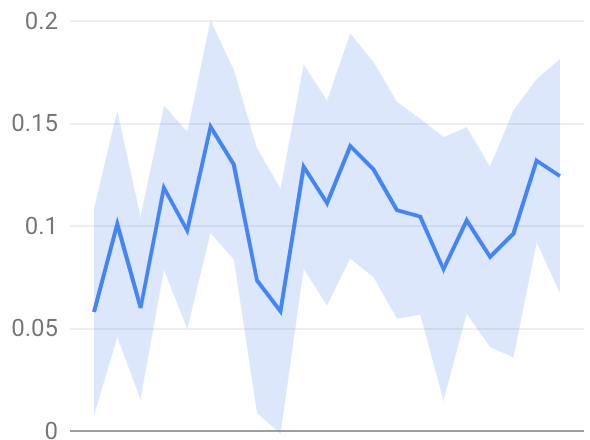}
        \caption{Diversity.}
    \end{subfigure}
    \caption{Live experiment results. On the x-axis is the date; on the y-axis is the relative difference in percentage between the experiment and control.}
    \label{fig:live_exp}
\end{figure}

We conduct A/B experiments in a live system serving billions of users to measure the benefits of the proposed latent intent modeling technique. 
The sequential recommender system~\citep{chen2019top} is built to retrieve hundreds of candidates from a corpus of millions of items upon each user request. The retrieved candidates, along with those returned by other sources, are scored and ranked by a separate ranking system before showing the top results to the user. 

Experiments are run for three weeks, during which both the control and experiment models are trained continuously with new interactions and feedback being used as training data. 
We focus on the following two metrics: 
(1) users’ overall enjoyment of the platform; 
(2) diversity of user-item interactions, which represents the number of unique topic clusters users have interacted with.
It has been shown that consumption diversity is an effective surrogate for long-term user experience~\citep{wang2022surrogate}.

The experiment and control models are the sequential recommender models with and without the latent intent module.
Figure~\ref{fig:live_exp} summarizes the live experiment results. 
On the x-axis is the date, and on the y-axis is the relative difference of a metric in percentage between the experiment and control.
We report the mean and 95\% confidence intervals of the metrics.
Relative to the control, the experiment model improves the overall enjoyment by $+0.07\%$ with a 95\% confidence interval of $(+0.02\%, +0.11\%)$.
Diversity of user-item interactions improves by $+0.10\%$ with a 95\% confidence interval of $(+0.08\%, +0.13\%)$.
Furthermore, there is an upward trend in the overall enjoyment metric, suggesting a user learning effect, i.e., user states change in response to the recommendation policy.

The proposed model has been deployed to the production system for more than two weeks. The way it is deployed is described at the beginning of this section.


\section{Conclusion}

In this work, we propose the latent user intent model, a probabilistic modeling approach to capturing user intent and complementing existing sequential recommenders.
The variational inference technique we adopt is efficient and scalable, and techniques to stabilize and interpret the model are also studied.
Finally, the effectiveness of the proposed method is validated in large-scale live experiments.
One future research direction is to further improve the interpretability of latent user intent by designing a discrete latent space, using Gumbel softmax~\citep{jang2017categorical,maddison2017the} or vector quantized variational autoencoder (VQVAE)~\citep{van2017neural}.

\bibliographystyle{ACM-Reference-Format}
\bibliography{www_industry}


\begin{thebibliography}{43}


\ifx \showCODEN    \undefined \def \showCODEN     #1{\unskip}     \fi
\ifx \showDOI      \undefined \def \showDOI       #1{#1}\fi
\ifx \showISBNx    \undefined \def \showISBNx     #1{\unskip}     \fi
\ifx \showISBNxiii \undefined \def \showISBNxiii  #1{\unskip}     \fi
\ifx \showISSN     \undefined \def \showISSN      #1{\unskip}     \fi
\ifx \showLCCN     \undefined \def \showLCCN      #1{\unskip}     \fi
\ifx \shownote     \undefined \def \shownote      #1{#1}          \fi
\ifx \showarticletitle \undefined \def \showarticletitle #1{#1}   \fi
\ifx \showURL      \undefined \def \showURL       {\relax}        \fi
\providecommand\bibfield[2]{#2}
\providecommand\bibinfo[2]{#2}
\providecommand\natexlab[1]{#1}
\providecommand\showeprint[2][]{arXiv:#2}

\bibitem[\protect\citeauthoryear{Aigner, Hsiao, Kapteyn, and Wansbeek}{Aigner
  et~al\mbox{.}}{1984}]%
        {aigner1984latent}
\bibfield{author}{\bibinfo{person}{Dennis~J Aigner}, \bibinfo{person}{Cheng
  Hsiao}, \bibinfo{person}{Arie Kapteyn}, {and} \bibinfo{person}{Tom
  Wansbeek}.} \bibinfo{year}{1984}\natexlab{}.
\newblock \showarticletitle{Latent variable models in econometrics}.
\newblock \bibinfo{journal}{\emph{Handbook of econometrics}}
  \bibinfo{volume}{2} (\bibinfo{year}{1984}), \bibinfo{pages}{1321--1393}.
\newblock


\bibitem[\protect\citeauthoryear{Alemi, Fischer, Dillon, and Murphy}{Alemi
  et~al\mbox{.}}{2017}]%
        {alemi2017deep}
\bibfield{author}{\bibinfo{person}{Alexander~A. Alemi}, \bibinfo{person}{Ian
  Fischer}, \bibinfo{person}{Joshua~V. Dillon}, {and} \bibinfo{person}{Kevin
  Murphy}.} \bibinfo{year}{2017}\natexlab{}.
\newblock \showarticletitle{Deep Variational Information Bottleneck}. In
  \bibinfo{booktitle}{\emph{International Conference on Learning
  Representations}}.
\newblock


\bibitem[\protect\citeauthoryear{Beutel, Covington, Jain, Xu, Li, Gatto, and
  Chi}{Beutel et~al\mbox{.}}{2018}]%
        {beutel2018latent}
\bibfield{author}{\bibinfo{person}{Alex Beutel}, \bibinfo{person}{Paul
  Covington}, \bibinfo{person}{Sagar Jain}, \bibinfo{person}{Can Xu},
  \bibinfo{person}{Jia Li}, \bibinfo{person}{Vince Gatto}, {and}
  \bibinfo{person}{Ed~H Chi}.} \bibinfo{year}{2018}\natexlab{}.
\newblock \showarticletitle{Latent cross: Making use of context in recurrent
  recommender systems}. In \bibinfo{booktitle}{\emph{Proceedings of the
  Eleventh ACM International Conference on Web Search and Data Mining}}.
  \bibinfo{pages}{46--54}.
\newblock


\bibitem[\protect\citeauthoryear{Bishop}{Bishop}{1998}]%
        {bishop1998latent}
\bibfield{author}{\bibinfo{person}{Christopher~M Bishop}.}
  \bibinfo{year}{1998}\natexlab{}.
\newblock \showarticletitle{Latent variable models}.
\newblock In \bibinfo{booktitle}{\emph{Learning in graphical models}}.
  \bibinfo{publisher}{Springer}, \bibinfo{pages}{371--403}.
\newblock


\bibitem[\protect\citeauthoryear{Broder}{Broder}{2002}]%
        {broder2002taxonomy}
\bibfield{author}{\bibinfo{person}{Andrei Broder}.}
  \bibinfo{year}{2002}\natexlab{}.
\newblock \showarticletitle{A taxonomy of web search}. In
  \bibinfo{booktitle}{\emph{ACM Sigir forum}}, Vol.~\bibinfo{volume}{36}. ACM
  New York, NY, USA, \bibinfo{pages}{3--10}.
\newblock


\bibitem[\protect\citeauthoryear{Cen, Zhang, Zou, Zhou, Yang, and Tang}{Cen
  et~al\mbox{.}}{2020}]%
        {cen2020controllable}
\bibfield{author}{\bibinfo{person}{Yukuo Cen}, \bibinfo{person}{Jianwei Zhang},
  \bibinfo{person}{Xu Zou}, \bibinfo{person}{Chang Zhou},
  \bibinfo{person}{Hongxia Yang}, {and} \bibinfo{person}{Jie Tang}.}
  \bibinfo{year}{2020}\natexlab{}.
\newblock \showarticletitle{Controllable multi-interest framework for
  recommendation}. In \bibinfo{booktitle}{\emph{Proceedings of the 26th ACM
  SIGKDD International Conference on Knowledge Discovery \& Data Mining}}.
  \bibinfo{pages}{2942--2951}.
\newblock


\bibitem[\protect\citeauthoryear{Chen, Beutel, Covington, Jain, Belletti, and
  Chi}{Chen et~al\mbox{.}}{2019}]%
        {chen2019top}
\bibfield{author}{\bibinfo{person}{Minmin Chen}, \bibinfo{person}{Alex Beutel},
  \bibinfo{person}{Paul Covington}, \bibinfo{person}{Sagar Jain},
  \bibinfo{person}{Francois Belletti}, {and} \bibinfo{person}{Ed~H Chi}.}
  \bibinfo{year}{2019}\natexlab{}.
\newblock \showarticletitle{Top-k off-policy correction for a REINFORCE
  recommender system}. In \bibinfo{booktitle}{\emph{WSDM}}.
  \bibinfo{pages}{456--464}.
\newblock


\bibitem[\protect\citeauthoryear{Chen, Liu, Li, McAuley, and Xiong}{Chen
  et~al\mbox{.}}{2022}]%
        {chen2022intent}
\bibfield{author}{\bibinfo{person}{Yongjun Chen}, \bibinfo{person}{Zhiwei Liu},
  \bibinfo{person}{Jia Li}, \bibinfo{person}{Julian McAuley}, {and}
  \bibinfo{person}{Caiming Xiong}.} \bibinfo{year}{2022}\natexlab{}.
\newblock \showarticletitle{Intent Contrastive Learning for Sequential
  Recommendation}. In \bibinfo{booktitle}{\emph{Proceedings of the ACM Web
  Conference 2022}}. \bibinfo{pages}{2172--2182}.
\newblock


\bibitem[\protect\citeauthoryear{Chua, Calandra, McAllister, and Levine}{Chua
  et~al\mbox{.}}{2018}]%
        {chua2018deep}
\bibfield{author}{\bibinfo{person}{Kurtland Chua}, \bibinfo{person}{Roberto
  Calandra}, \bibinfo{person}{Rowan McAllister}, {and} \bibinfo{person}{Sergey
  Levine}.} \bibinfo{year}{2018}\natexlab{}.
\newblock \showarticletitle{Deep reinforcement learning in a handful of trials
  using probabilistic dynamics models}.
\newblock \bibinfo{journal}{\emph{Advances in neural information processing
  systems}}  \bibinfo{volume}{31} (\bibinfo{year}{2018}).
\newblock


\bibitem[\protect\citeauthoryear{Chung, Kastner, Dinh, Goel, Courville, and
  Bengio}{Chung et~al\mbox{.}}{2015}]%
        {chung2015recurrent}
\bibfield{author}{\bibinfo{person}{Junyoung Chung}, \bibinfo{person}{Kyle
  Kastner}, \bibinfo{person}{Laurent Dinh}, \bibinfo{person}{Kratarth Goel},
  \bibinfo{person}{Aaron~C Courville}, {and} \bibinfo{person}{Yoshua Bengio}.}
  \bibinfo{year}{2015}\natexlab{}.
\newblock \showarticletitle{A recurrent latent variable model for sequential
  data}.
\newblock \bibinfo{journal}{\emph{Advances in neural information processing
  systems}}  \bibinfo{volume}{28} (\bibinfo{year}{2015}).
\newblock


\bibitem[\protect\citeauthoryear{Everett}{Everett}{2013}]%
        {everett2013introduction}
\bibfield{author}{\bibinfo{person}{B Everett}.}
  \bibinfo{year}{2013}\natexlab{}.
\newblock \bibinfo{booktitle}{\emph{An introduction to latent variable
  models}}.
\newblock \bibinfo{publisher}{Springer Science \& Business Media}.
\newblock


\bibitem[\protect\citeauthoryear{Friston}{Friston}{2010}]%
        {friston2010free}
\bibfield{author}{\bibinfo{person}{Karl Friston}.}
  \bibinfo{year}{2010}\natexlab{}.
\newblock \showarticletitle{The free-energy principle: a unified brain theory?}
\newblock \bibinfo{journal}{\emph{Nature reviews neuroscience}}
  \bibinfo{volume}{11}, \bibinfo{number}{2} (\bibinfo{year}{2010}),
  \bibinfo{pages}{127--138}.
\newblock


\bibitem[\protect\citeauthoryear{Hidasi, Karatzoglou, Baltrunas, and
  Tikk}{Hidasi et~al\mbox{.}}{2016}]%
        {Hidasi2016}
\bibfield{author}{\bibinfo{person}{Bal{\'a}zs Hidasi},
  \bibinfo{person}{Alexandros Karatzoglou}, \bibinfo{person}{Linas Baltrunas},
  {and} \bibinfo{person}{Domonkos Tikk}.} \bibinfo{year}{2016}\natexlab{}.
\newblock \showarticletitle{Session-based Recommendations with Recurrent Neural
  Networks}.
\newblock \bibinfo{journal}{\emph{International Conference of Learning
  Representations ICLR}} (\bibinfo{year}{2016}).
\newblock


\bibitem[\protect\citeauthoryear{Itti and Baldi}{Itti and Baldi}{2005}]%
        {itti2005bayesian}
\bibfield{author}{\bibinfo{person}{Laurent Itti} {and} \bibinfo{person}{Pierre
  Baldi}.} \bibinfo{year}{2005}\natexlab{}.
\newblock \showarticletitle{Bayesian surprise attracts human attention}.
\newblock \bibinfo{journal}{\emph{Advances in neural information processing
  systems}}  \bibinfo{volume}{18} (\bibinfo{year}{2005}).
\newblock


\bibitem[\protect\citeauthoryear{Jang, Gu, and Poole}{Jang
  et~al\mbox{.}}{2017}]%
        {jang2017categorical}
\bibfield{author}{\bibinfo{person}{Eric Jang}, \bibinfo{person}{Shixiang Gu},
  {and} \bibinfo{person}{Ben Poole}.} \bibinfo{year}{2017}\natexlab{}.
\newblock \showarticletitle{Categorical Reparameterization with
  Gumbel-Softmax}. In \bibinfo{booktitle}{\emph{International Conference on
  Learning Representations}}.
\newblock


\bibitem[\protect\citeauthoryear{Jansen, Booth, and Spink}{Jansen
  et~al\mbox{.}}{2007}]%
        {jansen2007determining}
\bibfield{author}{\bibinfo{person}{Bernard~J Jansen},
  \bibinfo{person}{Danielle~L Booth}, {and} \bibinfo{person}{Amanda Spink}.}
  \bibinfo{year}{2007}\natexlab{}.
\newblock \showarticletitle{Determining the user intent of web search engine
  queries}. In \bibinfo{booktitle}{\emph{Proceedings of the 16th international
  conference on World Wide Web}}. \bibinfo{pages}{1149--1150}.
\newblock


\bibitem[\protect\citeauthoryear{Kalaganis, Georgiadis, Oikonomou, Laskaris,
  Nikolopoulos, and Kompatsiaris}{Kalaganis et~al\mbox{.}}{2021}]%
        {kalaganis2021unlocking}
\bibfield{author}{\bibinfo{person}{Fotis~P Kalaganis}, \bibinfo{person}{Kostas
  Georgiadis}, \bibinfo{person}{Vangelis~P Oikonomou}, \bibinfo{person}{Nikos~A
  Laskaris}, \bibinfo{person}{Spiros Nikolopoulos}, {and}
  \bibinfo{person}{Ioannis Kompatsiaris}.} \bibinfo{year}{2021}\natexlab{}.
\newblock \showarticletitle{Unlocking the Subconscious Consumer Bias: A Survey
  on the Past, Present, and Future of Hybrid EEG Schemes in Neuromarketing}.
\newblock \bibinfo{journal}{\emph{Frontiers in Neuroergonomics}}
  \bibinfo{volume}{2} (\bibinfo{year}{2021}), \bibinfo{pages}{672982}.
\newblock


\bibitem[\protect\citeauthoryear{Kang and McAuley}{Kang and McAuley}{2018}]%
        {kang2018self}
\bibfield{author}{\bibinfo{person}{Wang-Cheng Kang} {and}
  \bibinfo{person}{Julian McAuley}.} \bibinfo{year}{2018}\natexlab{}.
\newblock \showarticletitle{Self-attentive sequential recommendation}. In
  \bibinfo{booktitle}{\emph{2018 IEEE international conference on data mining
  (ICDM)}}. IEEE, \bibinfo{pages}{197--206}.
\newblock


\bibitem[\protect\citeauthoryear{Kingma and Welling}{Kingma and
  Welling}{2013}]%
        {kingma2013auto}
\bibfield{author}{\bibinfo{person}{Diederik~P Kingma} {and}
  \bibinfo{person}{Max Welling}.} \bibinfo{year}{2013}\natexlab{}.
\newblock \showarticletitle{Auto-encoding variational bayes}.
\newblock \bibinfo{journal}{\emph{arXiv preprint arXiv:1312.6114}}
  (\bibinfo{year}{2013}).
\newblock


\bibitem[\protect\citeauthoryear{Kingma and Welling}{Kingma and
  Welling}{2019}]%
        {Kingma2019}
\bibfield{author}{\bibinfo{person}{Diederik~P. Kingma} {and}
  \bibinfo{person}{Max Welling}.} \bibinfo{year}{2019}\natexlab{}.
\newblock \showarticletitle{An Introduction to Variational Autoencoders}.
\newblock \bibinfo{journal}{\emph{Foundations and Trends® in Machine
  Learning}} \bibinfo{volume}{12}, \bibinfo{number}{4} (\bibinfo{year}{2019}),
  \bibinfo{pages}{307--392}.
\newblock
\showISSN{1935-8237}


\bibitem[\protect\citeauthoryear{Li, Liu, Wu, Xu, Zhao, Huang, Kang, Chen, Li,
  and Lee}{Li et~al\mbox{.}}{2019}]%
        {li2019multi}
\bibfield{author}{\bibinfo{person}{Chao Li}, \bibinfo{person}{Zhiyuan Liu},
  \bibinfo{person}{Mengmeng Wu}, \bibinfo{person}{Yuchi Xu},
  \bibinfo{person}{Huan Zhao}, \bibinfo{person}{Pipei Huang},
  \bibinfo{person}{Guoliang Kang}, \bibinfo{person}{Qiwei Chen},
  \bibinfo{person}{Wei Li}, {and} \bibinfo{person}{Dik~Lun Lee}.}
  \bibinfo{year}{2019}\natexlab{}.
\newblock \showarticletitle{Multi-interest network with dynamic routing for
  recommendation at Tmall}. In \bibinfo{booktitle}{\emph{Proceedings of the
  28th ACM International Conference on Information and Knowledge Management}}.
  \bibinfo{pages}{2615--2623}.
\newblock


\bibitem[\protect\citeauthoryear{Li, Wang, Zhang, Ma, Cui, and Zhu}{Li
  et~al\mbox{.}}{2021}]%
        {li2021intention}
\bibfield{author}{\bibinfo{person}{Haoyang Li}, \bibinfo{person}{Xin Wang},
  \bibinfo{person}{Ziwei Zhang}, \bibinfo{person}{Jianxin Ma},
  \bibinfo{person}{Peng Cui}, {and} \bibinfo{person}{Wenwu Zhu}.}
  \bibinfo{year}{2021}\natexlab{}.
\newblock \showarticletitle{Intention-aware sequential recommendation with
  structured intent transition}.
\newblock \bibinfo{journal}{\emph{IEEE Transactions on Knowledge and Data
  Engineering}} (\bibinfo{year}{2021}).
\newblock


\bibitem[\protect\citeauthoryear{Li and She}{Li and She}{2017}]%
        {li2017collaborative}
\bibfield{author}{\bibinfo{person}{Xiaopeng Li} {and} \bibinfo{person}{James
  She}.} \bibinfo{year}{2017}\natexlab{}.
\newblock \showarticletitle{Collaborative variational autoencoder for
  recommender systems}. In \bibinfo{booktitle}{\emph{Proceedings of the 23rd
  ACM SIGKDD international conference on knowledge discovery and data mining}}.
  \bibinfo{pages}{305--314}.
\newblock


\bibitem[\protect\citeauthoryear{Li, Gao, Du, Wei, Luo, Jin, and Li}{Li
  et~al\mbox{.}}{2022}]%
        {li2022automatically}
\bibfield{author}{\bibinfo{person}{Yinfeng Li}, \bibinfo{person}{Chen Gao},
  \bibinfo{person}{Xiaoyi Du}, \bibinfo{person}{Huazhou Wei},
  \bibinfo{person}{Hengliang Luo}, \bibinfo{person}{Depeng Jin}, {and}
  \bibinfo{person}{Yong Li}.} \bibinfo{year}{2022}\natexlab{}.
\newblock \showarticletitle{Automatically Discovering User Consumption Intents
  in Meituan}. In \bibinfo{booktitle}{\emph{Proceedings of the 28th ACM SIGKDD
  Conference on Knowledge Discovery and Data Mining}}.
  \bibinfo{pages}{3259--3269}.
\newblock


\bibitem[\protect\citeauthoryear{Liang, Krishnan, Hoffman, and Jebara}{Liang
  et~al\mbox{.}}{2018}]%
        {liang2018variational}
\bibfield{author}{\bibinfo{person}{Dawen Liang}, \bibinfo{person}{Rahul~G
  Krishnan}, \bibinfo{person}{Matthew~D Hoffman}, {and} \bibinfo{person}{Tony
  Jebara}.} \bibinfo{year}{2018}\natexlab{}.
\newblock \showarticletitle{Variational autoencoders for collaborative
  filtering}. In \bibinfo{booktitle}{\emph{Proceedings of the 2018 world wide
  web conference}}. \bibinfo{pages}{689--698}.
\newblock


\bibitem[\protect\citeauthoryear{Luo, Yang, Wu, and Sanner}{Luo
  et~al\mbox{.}}{2020}]%
        {luo2020deep}
\bibfield{author}{\bibinfo{person}{Kai Luo}, \bibinfo{person}{Hojin Yang},
  \bibinfo{person}{Ga Wu}, {and} \bibinfo{person}{Scott Sanner}.}
  \bibinfo{year}{2020}\natexlab{}.
\newblock \showarticletitle{Deep critiquing for VAE-based recommender systems}.
  In \bibinfo{booktitle}{\emph{Proceedings of the 43rd International ACM SIGIR
  Conference on Research and Development in Information Retrieval}}.
  \bibinfo{pages}{1269--1278}.
\newblock


\bibitem[\protect\citeauthoryear{Ma, Gong, Hern{\'a}ndez-Lobato, Koenigstein,
  Nowozin, and Zhang}{Ma et~al\mbox{.}}{2018}]%
        {ma2018partial}
\bibfield{author}{\bibinfo{person}{Chao Ma}, \bibinfo{person}{Wenbo Gong},
  \bibinfo{person}{Jos{\'e}~Miguel Hern{\'a}ndez-Lobato}, \bibinfo{person}{Noam
  Koenigstein}, \bibinfo{person}{Sebastian Nowozin}, {and}
  \bibinfo{person}{Cheng Zhang}.} \bibinfo{year}{2018}\natexlab{}.
\newblock \showarticletitle{Partial VAE for hybrid recommender system}. In
  \bibinfo{booktitle}{\emph{NIPS Workshop on Bayesian Deep Learning}},
  Vol.~\bibinfo{volume}{2018}.
\newblock


\bibitem[\protect\citeauthoryear{Ma, Zhou, Yang, Cui, Wang, and Zhu}{Ma
  et~al\mbox{.}}{2020}]%
        {ma2020disentangled}
\bibfield{author}{\bibinfo{person}{Jianxin Ma}, \bibinfo{person}{Chang Zhou},
  \bibinfo{person}{Hongxia Yang}, \bibinfo{person}{Peng Cui},
  \bibinfo{person}{Xin Wang}, {and} \bibinfo{person}{Wenwu Zhu}.}
  \bibinfo{year}{2020}\natexlab{}.
\newblock \showarticletitle{Disentangled self-supervision in sequential
  recommenders}. In \bibinfo{booktitle}{\emph{Proceedings of the 26th ACM
  SIGKDD International Conference on Knowledge Discovery \& Data Mining}}.
  \bibinfo{pages}{483--491}.
\newblock


\bibitem[\protect\citeauthoryear{Maddison, Mnih, and Teh}{Maddison
  et~al\mbox{.}}{2017}]%
        {maddison2017the}
\bibfield{author}{\bibinfo{person}{Chris~J. Maddison}, \bibinfo{person}{Andriy
  Mnih}, {and} \bibinfo{person}{Yee~Whye Teh}.}
  \bibinfo{year}{2017}\natexlab{}.
\newblock \showarticletitle{The Concrete Distribution: A Continuous Relaxation
  of Discrete Random Variables}. In \bibinfo{booktitle}{\emph{International
  Conference on Learning Representations}}.
\newblock


\bibitem[\protect\citeauthoryear{Mehrotra, Lalmas, Kenney, Lim-Meng, and
  Hashemian}{Mehrotra et~al\mbox{.}}{2019}]%
        {spotify}
\bibfield{author}{\bibinfo{person}{Rishabh Mehrotra}, \bibinfo{person}{Mounia
  Lalmas}, \bibinfo{person}{Doug Kenney}, \bibinfo{person}{Thomas Lim-Meng},
  {and} \bibinfo{person}{Golli Hashemian}.} \bibinfo{year}{2019}\natexlab{}.
\newblock \showarticletitle{Jointly leveraging intent and interaction signals
  to predict user satisfaction with slate recommendations}. In
  \bibinfo{booktitle}{\emph{The World Wide Web Conference}}.
  \bibinfo{pages}{1256--1267}.
\newblock


\bibitem[\protect\citeauthoryear{Pan, Cai, Ling, and de~Rijke}{Pan
  et~al\mbox{.}}{2020}]%
        {pan2020intent}
\bibfield{author}{\bibinfo{person}{Zhiqiang Pan}, \bibinfo{person}{Fei Cai},
  \bibinfo{person}{Yanxiang Ling}, {and} \bibinfo{person}{Maarten de Rijke}.}
  \bibinfo{year}{2020}\natexlab{}.
\newblock \showarticletitle{An intent-guided collaborative machine for
  session-based recommendation}. In \bibinfo{booktitle}{\emph{Proceedings of
  the 43rd international ACM SIGIR conference on research and development in
  information retrieval}}. \bibinfo{pages}{1833--1836}.
\newblock


\bibitem[\protect\citeauthoryear{Rose and Levinson}{Rose and Levinson}{2004}]%
        {rose2004understanding}
\bibfield{author}{\bibinfo{person}{Daniel~E Rose} {and} \bibinfo{person}{Danny
  Levinson}.} \bibinfo{year}{2004}\natexlab{}.
\newblock \showarticletitle{Understanding user goals in web search}. In
  \bibinfo{booktitle}{\emph{Proceedings of the 13th international conference on
  World Wide Web}}. \bibinfo{pages}{13--19}.
\newblock


\bibitem[\protect\citeauthoryear{Sohn, Lee, and Yan}{Sohn
  et~al\mbox{.}}{2015}]%
        {sohn2015learning}
\bibfield{author}{\bibinfo{person}{Kihyuk Sohn}, \bibinfo{person}{Honglak Lee},
  {and} \bibinfo{person}{Xinchen Yan}.} \bibinfo{year}{2015}\natexlab{}.
\newblock \showarticletitle{Learning structured output representation using
  deep conditional generative models}.
\newblock \bibinfo{journal}{\emph{Advances in neural information processing
  systems}}  \bibinfo{volume}{28} (\bibinfo{year}{2015}).
\newblock


\bibitem[\protect\citeauthoryear{Tang, Belletti, Jain, Chen, Beutel, Xu, and
  H.~Chi}{Tang et~al\mbox{.}}{2019}]%
        {tang2019towards}
\bibfield{author}{\bibinfo{person}{Jiaxi Tang}, \bibinfo{person}{Francois
  Belletti}, \bibinfo{person}{Sagar Jain}, \bibinfo{person}{Minmin Chen},
  \bibinfo{person}{Alex Beutel}, \bibinfo{person}{Can Xu}, {and}
  \bibinfo{person}{Ed H.~Chi}.} \bibinfo{year}{2019}\natexlab{}.
\newblock \showarticletitle{Towards neural mixture recommender for long range
  dependent user sequences}. In \bibinfo{booktitle}{\emph{The World Wide Web
  Conference}}. \bibinfo{pages}{1782--1793}.
\newblock


\bibitem[\protect\citeauthoryear{Tang and Wang}{Tang and Wang}{2018}]%
        {tang2018personalized}
\bibfield{author}{\bibinfo{person}{Jiaxi Tang} {and} \bibinfo{person}{Ke
  Wang}.} \bibinfo{year}{2018}\natexlab{}.
\newblock \showarticletitle{Personalized top-n sequential recommendation via
  convolutional sequence embedding}. In \bibinfo{booktitle}{\emph{Proceedings
  of the eleventh ACM international conference on web search and data mining}}.
  \bibinfo{pages}{565--573}.
\newblock


\bibitem[\protect\citeauthoryear{Tanjim, Su, Benjamin, Hu, Hong, and
  McAuley}{Tanjim et~al\mbox{.}}{2020}]%
        {tanjim2020attentive}
\bibfield{author}{\bibinfo{person}{Md~Mehrab Tanjim}, \bibinfo{person}{Congzhe
  Su}, \bibinfo{person}{Ethan Benjamin}, \bibinfo{person}{Diane Hu},
  \bibinfo{person}{Liangjie Hong}, {and} \bibinfo{person}{Julian McAuley}.}
  \bibinfo{year}{2020}\natexlab{}.
\newblock \showarticletitle{Attentive sequential models of latent intent for
  next item recommendation}. In \bibinfo{booktitle}{\emph{Proceedings of The
  Web Conference 2020}}. \bibinfo{pages}{2528--2534}.
\newblock


\bibitem[\protect\citeauthoryear{Van Den~Oord, Vinyals, et~al\mbox{.}}{Van
  Den~Oord et~al\mbox{.}}{2017}]%
        {van2017neural}
\bibfield{author}{\bibinfo{person}{Aaron Van Den~Oord}, \bibinfo{person}{Oriol
  Vinyals}, {et~al\mbox{.}}} \bibinfo{year}{2017}\natexlab{}.
\newblock \showarticletitle{Neural discrete representation learning}.
\newblock \bibinfo{journal}{\emph{Advances in neural information processing
  systems}}  \bibinfo{volume}{30} (\bibinfo{year}{2017}).
\newblock


\bibitem[\protect\citeauthoryear{Wang, Hu, Wang, Cao, Sheng, and Orgun}{Wang
  et~al\mbox{.}}{2019a}]%
        {ijcai2019p883}
\bibfield{author}{\bibinfo{person}{Shoujin Wang}, \bibinfo{person}{Liang Hu},
  \bibinfo{person}{Yan Wang}, \bibinfo{person}{Longbing Cao},
  \bibinfo{person}{Quan~Z. Sheng}, {and} \bibinfo{person}{Mehmet Orgun}.}
  \bibinfo{year}{2019}\natexlab{a}.
\newblock \showarticletitle{Sequential Recommender Systems: Challenges,
  Progress and Prospects}. In \bibinfo{booktitle}{\emph{Proceedings of the
  Twenty-Eighth International Joint Conference on Artificial Intelligence,
  {IJCAI-19}}}. \bibinfo{pages}{6332--6338}.
\newblock


\bibitem[\protect\citeauthoryear{Wang, Hu, Wang, Sheng, Orgun, and Cao}{Wang
  et~al\mbox{.}}{2019b}]%
        {wang2019modeling}
\bibfield{author}{\bibinfo{person}{Shoujin Wang}, \bibinfo{person}{Liang Hu},
  \bibinfo{person}{Yan Wang}, \bibinfo{person}{Quan~Z Sheng},
  \bibinfo{person}{Mehmet Orgun}, {and} \bibinfo{person}{Longbing Cao}.}
  \bibinfo{year}{2019}\natexlab{b}.
\newblock \showarticletitle{Modeling multi-purpose sessions for next-item
  recommendations via mixture-channel purpose routing networks}. In
  \bibinfo{booktitle}{\emph{International Joint Conference on Artificial
  Intelligence}}. International Joint Conferences on Artificial Intelligence.
\newblock


\bibitem[\protect\citeauthoryear{Wang, Sharma, Xu, Badam, Sun, Richardson,
  Chung, Chi, and Chen}{Wang et~al\mbox{.}}{2022}]%
        {wang2022surrogate}
\bibfield{author}{\bibinfo{person}{Yuyan Wang}, \bibinfo{person}{Mohit Sharma},
  \bibinfo{person}{Can Xu}, \bibinfo{person}{Sriraj Badam},
  \bibinfo{person}{Qian Sun}, \bibinfo{person}{Lee Richardson},
  \bibinfo{person}{Lisa Chung}, \bibinfo{person}{Ed~H Chi}, {and}
  \bibinfo{person}{Minmin Chen}.} \bibinfo{year}{2022}\natexlab{}.
\newblock \showarticletitle{Surrogate for Long-Term User Experience in
  Recommender Systems}. In \bibinfo{booktitle}{\emph{Proceedings of the 28th
  ACM SIGKDD Conference on Knowledge Discovery and Data Mining}}.
  \bibinfo{pages}{4100--4109}.
\newblock


\bibitem[\protect\citeauthoryear{Williams}{Williams}{1992}]%
        {williams1992simple}
\bibfield{author}{\bibinfo{person}{Ronald~J Williams}.}
  \bibinfo{year}{1992}\natexlab{}.
\newblock \showarticletitle{Simple statistical gradient-following algorithms
  for connectionist reinforcement learning}.
\newblock \bibinfo{journal}{\emph{Machine learning}} \bibinfo{volume}{8},
  \bibinfo{number}{3-4} (\bibinfo{year}{1992}), \bibinfo{pages}{229--256}.
\newblock


\bibitem[\protect\citeauthoryear{Wu, Ahmed, Beutel, Smola, and Jing}{Wu
  et~al\mbox{.}}{2017}]%
        {wu2017recurrent}
\bibfield{author}{\bibinfo{person}{Chao-Yuan Wu}, \bibinfo{person}{Amr Ahmed},
  \bibinfo{person}{Alex Beutel}, \bibinfo{person}{Alexander~J Smola}, {and}
  \bibinfo{person}{How Jing}.} \bibinfo{year}{2017}\natexlab{}.
\newblock \showarticletitle{Recurrent recommender networks}. In
  \bibinfo{booktitle}{\emph{Proceedings of the tenth ACM international
  conference on web search and data mining}}. \bibinfo{pages}{495--503}.
\newblock


\bibitem[\protect\citeauthoryear{Yuan, Karatzoglou, Arapakis, Jose, and
  He}{Yuan et~al\mbox{.}}{2019}]%
        {yuan2019simple}
\bibfield{author}{\bibinfo{person}{Fajie Yuan}, \bibinfo{person}{Alexandros
  Karatzoglou}, \bibinfo{person}{Ioannis Arapakis}, \bibinfo{person}{Joemon~M
  Jose}, {and} \bibinfo{person}{Xiangnan He}.} \bibinfo{year}{2019}\natexlab{}.
\newblock \showarticletitle{A simple convolutional generative network for next
  item recommendation}. In \bibinfo{booktitle}{\emph{Proceedings of the twelfth
  ACM international conference on web search and data mining}}.
  \bibinfo{pages}{582--590}.
\newblock


\end{thebibliography}

\end{document}